\documentclass[prc,twocolumn,showpacs,preprintnumbers,floatfix]{revtex4-1}
\pdfoutput=1
\usepackage{amsfonts,amsmath,amssymb}
\usepackage{graphicx}
\usepackage{color}
\usepackage{lipsum}

\begin{document}

\title{Search for Variations of $^{213}$Po Half-Life}

\newcommand{\BNOINR}{Baksan Neutrino Observatory INR RAS, Neitrino 361609, Russia}
\newcommand{\KhNU}{V.N.~Karazin Kharkiv National University, Kharkiv 61022, Ukraine}

\affiliation{\BNOINR}
\affiliation{\KhNU}

\author{E.N.~Alexeyev} \affiliation{\BNOINR}
\author{Yu.M.~Gavrilyuk} \affiliation{\BNOINR}
\author{A.M.~Gangapshev} \affiliation{\BNOINR}
\author{A.M.~Gezhaev} \affiliation{\BNOINR}
\author{V.V.~Kazalov} \affiliation{\BNOINR}
\author{V.V.~Kuzminov} \affiliation{\BNOINR}
\author{S.I.~Panasenko} \affiliation{\KhNU}
\author{S.S.~Ratkevich} \affiliation{\KhNU}


\begin{abstract}

A device with the parent $^{229}$Th source was constructed to search for variations of the daughter $^{213}$Po
half-life ($T_{1/2} = 4.2$ $\mu$s). A solar-daily variation with amplitude $A_{So}=(5.3 \pm 1.1) \times 10^{-4}$, a lunar-daily variation
with amplitude $A_L = (4.8 \pm 2.1) \times 10^{-4}$, and a sidereal-daily variation with amplitude $A_S = (4.2 \pm 1.7) \times 10^{-4}$ were found upon proceeding the data series over a 622-day interval (from July 2015 to March 2017). The
$^{213}$Po half-life mean value is found to be $T_{1/2} = 3.705 \pm  0.001$ $\mu$s.
The obtained half-life is in good agreement with some of the literature values obtained with great accuracy.
\end{abstract}
\maketitle

\section{\label{sec:intro}Introduction}
Experimental studies of $^{214}$Po half-life time dependence
$(\tau)$ \cite{a1,a2,a3} have been carried out at the Baksan
Neutrino Observatory of the Institute for Nuclear
Research of the Russian Academy of Sciences since
2008. Unlike studies on determining the half-life by
the results of analysis of time dependence of the
explored isotope activity, decay curves analyzed at the
Observatory are plotted based on a set of data on lifetimes
of separate nuclei of the $^{214}$Po isotope. In order
to determine this parameter, the delays between the
moment of a nucleus production (a beta electron from
$^{214}$Bi decay + a gamma-quantum) and its decay (an
alpha particle from the $^{214}$Po decay) are measured. The
measurements are performed at TAU-2, a low-background
facility placed in the underground low-background
laboratory NLGZ-4900 at the depth of
4900 m.w.e. (973 days) and at TAU-1 of the underground
low-background KAPRIZ laboratory at the
depth of 1000 m.w.e. (354 days). The time interval of
measurements at TAU-1 corresponds to the end of the
measuring interval at TAU-2. Further, time series of $\tau$
with different temporal steps are analyzed. According
to data obtained at TAU-2, the averaged value of the
$^{214}$Po half-life is $\tau = 163.47 \pm 0.03$ $\mu$s. The annual variation
with amplitude $A = (9.8 \pm 0.6) \times 10^{-4}$, the solar-daily
variation with amplitude $A_{So} = (7.5 \pm 1.2) \times 10^{-4}$,
the lunar-daily variation with amplitude $A_L = (6.9 \pm 2.0) \times 10^{-4}$, and the sidereal-daily variation with
amplitude $A_S = (7.2 \pm 1.2) \times 10^{-4}$ are detected in the
series of $\tau$ values.

It was found that $\tau$-amplitude maxima are
observed at the moments when the maximum projection
is reached by the Earth’s surface point velocity
vector directed at the explored source of possible variations (the Sun, the Moon, or an unidentified stellar
object). Basically, it would be possible to explain the
origin of solar and lunar variations by these objects’
influence on performances of measuring facilities
through cyclical geophysical and climatic disturbances
(tidal waves, meteorological factors, magnetic field,
etc.) caused by them on the Earth. However, no reasonable
fundamental process capable of transforming
these disturbances into variations of time parameters
of measuring facilities in the required phase has been
yet discovered.

\begin{figure*}[pt]
\includegraphics[width=3.05in,angle=0.]{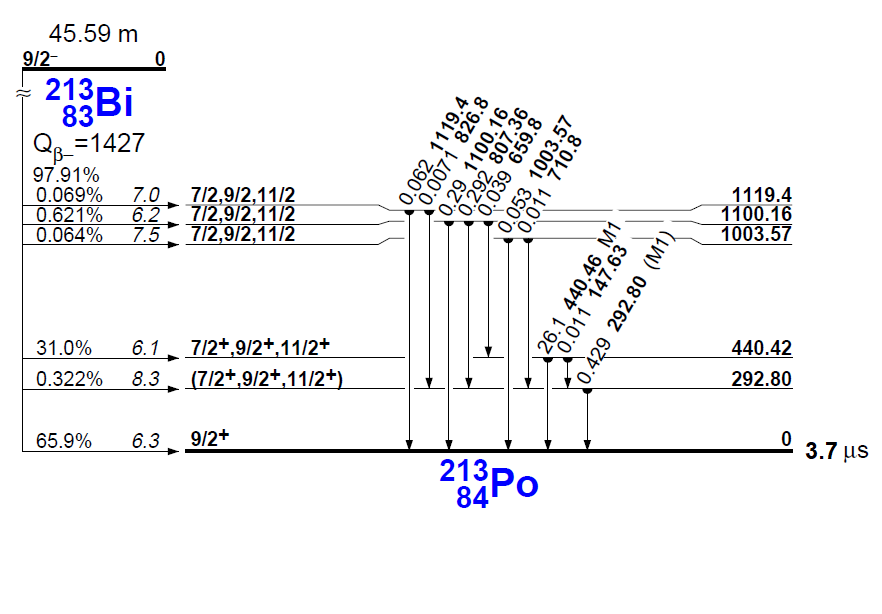}%
\includegraphics[width=2.55in,angle=0.]{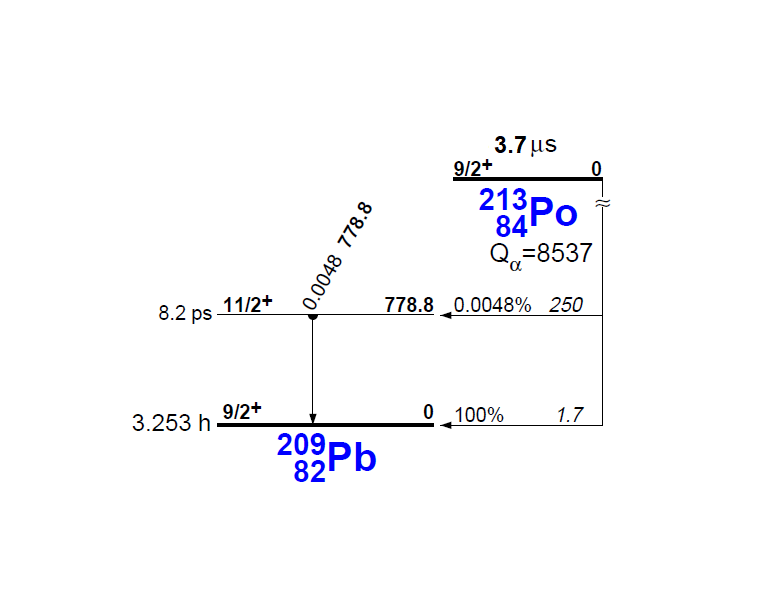}%
\caption{\label{f1}
Decay schemes of $^{213}$Bi and $^{213}$Po. }
\end{figure*}

At the same time, the sidereal-daily variation of $\tau$
observed, if this is not an instrumental effect, may
identify the presence of a real unknown physical phenomenon
influencing the parameter under study. Two
checks have been performed to verify the reliability of
the results. The dependencies of the amplitude and
phase of the observed sidereal-daily wave on the
choice of starting point of the analyzed time series
were verified. The start was shifted by 91 days and
182 days. The previous series section was not considered
in the analysis. As expected for this type of variation,
waves were obtained with amplitudes coinciding
with the initial values within the error, but with phases
shifted by 6 and 12 hours relative to the initial phases.
The second verification was performed by plotting the
daily data set in an anti-sidereal time (nonexistent periodicity).
The day duration in the anti-sidereal time was
increased relative to solar day duration by an interval
shortening the day duration in sidereal time. In such a
data set, within the statistical error $\pm 1.2 \times 10^{-4}$, the wave was absent.

Measurements were carried out at the TAU-1 facility
to confirm the non-random character of the
observed variations of the $\tau$ temporal series. Data provided
by TAU-1 revealed a solar-daily variation with amplitude $A_{So} = (17 \pm 3) \times 10^{-4}$, a lunar-daily variation
with amplitude $A_L = (8 \pm 3) \pm 10^{-4}$, and a sidereal-
daily variation with amplitude $A_S = (11 \pm 4) \times 10^{-4}$.
Searching for annual variations, initial data for a half
year were summed and then sequentially shifted to
increase the statistical reliability of the decay curve. As
a result, a $\tau$ series of no more than six-month duration
was obtained from the annual data set. Hence, it was
impossible to identify the annual periodicity with adequate
accuracy, so it was not explored. It is clear that a
substantial improvement in the statistical error can be
achieved by a substantial increase in the data-taking
rate. However, in the case of the $^{214}$Po isotope, the
speedup of data-taking induces a quadratic increase in
the share of random coincidences, up to $\sim1$\%
at a rate of 12 s$^{-1}$. This is caused by a high aggregate activity of all daughter $^{226}$Ra isotopes and the relatively long half-life
of $^{214}$Po ($\sim163.5$ $\mu$s).
Therefore, without augmentation
of the relative contribution of the random coincidence
background, the increase in the statistics
accumulation rate for $^{214}$Po can be achieved only by
the increase in the number of independent setups.
This might appear technologically infeasible. Another
alternative is to use a pair of radioactive isotopes having
a similar decay scheme, but essentially shorter half-life
of the daughter isotope. The following isotopes
might be used as such pairs: $^{213}$Bi ($T_{1/2} = 46$ min) $\rightarrow$ $^{213}$Po (weighted average of $T_{1/2} = 3.72(2)$ $\mu$s) \cite{a4a}
that are daughter products in the series of decays $^{229}$Th ($T_{1/2} = 7340$ years) from the
series of $^{237}$Np decays \cite{a5}. In this paper, the first results
obtained from using the facility with the specified
source are presented.

\section{THE FACILITY DESCRIPTION}

\begin{figure}
\includegraphics[width=2.8in,angle=0.]{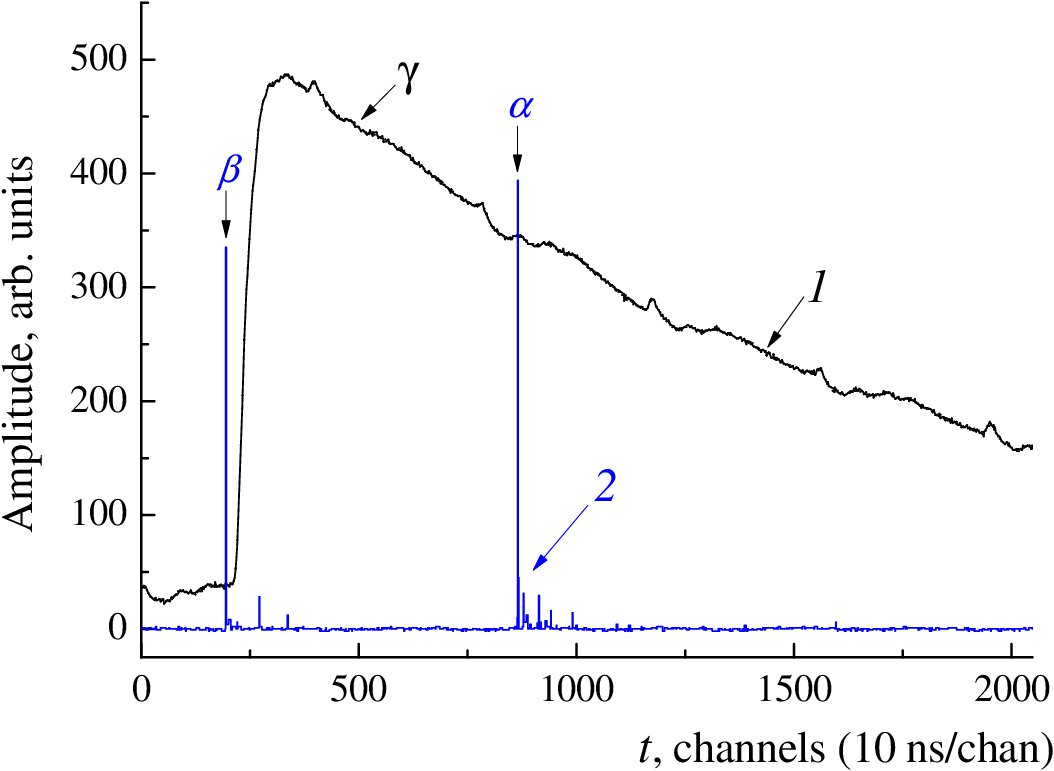}%
\caption{\label{f2} An example of $^{213}$Bi--$^{213}$Po pair decay event stored
by DO in PC memory: (\emph{1}) upper beam, a pulse from D2
detector ($\gamma$ quantum), (\emph{2}) lower beam, pulses from β particle
(start), and α particle (stop) in D1 detector.
}
\end{figure}

The construction of the TAU-3 facility with a $^{229}$Th
source is similar to TAU-1 and TAU-2 \cite{a1}. It comprises
a scintillation detector D1, a plastic scintillator
(PS), which is made of two disks $d = 18$ mm and $h = 1$ mm glued together. The radiation source $^{229}$Th
($T_{1/2} = 7340$ years) positioned between the disks is the
parent isotope for $^{213}$Po. The test sample is manufactured
at the Khlopin Radium Institute (St. Petersburg).
The source is precipitated from Th(NO$_3$)$_4$ salt
solution on the surface of a LAVSAN film with $h =
2.5$ $\mu$m and covered by the same film pasted along the
edge by the epoxy resin. The assembly is placed at the
bottom of a case made of VM-2000 reflecting film
open from one end. The case is put inside a stainless
steel rectangular case $9 \times 23 \times 140$ mm, thickness
0.5 mm. The open end of the case is connected with
the bottom of a 2.5-mm stainless steel cylinder with
$d = 44$ mm, and $h = 160$ mm. Inside the cylinder, there
is a high-speed FEU-87 photomultiplier monitoring
PS. The signal is taken from the FEU anode load
through the matching circuit and supplied via the
cable (50 Ohm) to the first entry of the registering
detector. Detector D1 is placed in the 15-cm Pb protective
layer in a gap with $h = 10$ mm between two scintillation
detectors NaI(Tl) $150 \times 150$ mm (detector D2)
in a low-background box of the NLGZ-4900 underground
low-background laboratory \cite{a6}. Signals from
the anodes of two photomultipliers of the D2 detector
are amplified by charge-sensitive preamplifiers,
summed, and supplied to the second, starting entry of
the registering detector. The registering facility comprises
a LA-n10-12 PCI digital oscilloscope (DO)
integrated with a PC that is registering in online mode
the waveform of pulses arriving from D1 and D2. The
frequency of pulse digitization in DO is chosen as
100 MHz. The reading and recording is started by a
pulse in the D2 channel. The record frame is
2048 temporal channels (10 ns per channel), including
256 channels of prehistory and 1792 channels of history.
In Fig.~\ref{f1}, the decays of $^{213}$Bi and $^{213}$Po isotopes
\cite{a4b} are presented schematically. From Fig.~\ref{f1}$a$ it
follows that 66\%
of $\beta$ decays of $^{213}$Bi are transitions to
the ground level, and 31\%
to the excited level with an
energy of 440 keV. The decay of this level is accompanied by a $\gamma$ quantum emission (26\%
per decay).
The isotope of $^{213}$Po decays in 100\%
of cases with emission
of an $\alpha$ particle with an energy of 8537 keV. If the
device registers all three particles released by the decay
of the pair of isotopes, it is the event with three pulses.
In this event, pulses coming from the $\gamma$ quantum and
$\beta$ particle coincide instantaneously, and the pulse from
the $\alpha$ is delayed. In Fig.~\ref{f2}, one of the events
(frames) stored by DO in the PC memory is displayed
as an example.
The pulse on the upper beam (\emph{1}) is a $\gamma$
quantum, the first pulse train on the lower beam (\emph{2})
corresponds to a $\beta$ particle, and the second one to an
$\alpha$ particle. The observed triple coincidences considerably
reduce the contribution of background events
accompanying decays of the remaining isotopes in the
chain of decays of $^{229}$Th to the total counting rate of
the facility. The activity of $^{229}$Th is $\sim80$ Bq. Alongside
the main isotope there are small amounts of extraneous
radioactive impurities in the specimen. The rate of
the event recording started by DO by D2 pulses with
amplitudes of 380–500 keV was $\sim$27 s$^{-1}$. The rate of
recording useful events with parameters of all pulses
corresponding to $^{213}$Po decay was $\sim18$ s$^{-1}$.

Following from Fig.~\ref{f2}, signals from β and α particles
are trains of short subpulses with total duration of
up to $\sim1$ $\mu$s, decreasing exponentially in frequency and
amplitude. The trains can overlap at small delays
between particles; therefore, the processing program
should consider relations between the amplitudes of
the first and subsequent subpulses in a train in order to
unambiguously separate the delayed $(\beta \otimes \alpha)$ coincidences.

The delays between pulses in channel D1 are determined
by the results of processing the recorded oscillograms,
and a decay curve of daughter isotope $^{213}$Po is plotted for the chosen time interval. The half-life
determination is based on this curve. The sequential
time series of this magnitude is plotted.

\section{MEASUREMENT RESULTS}

\begin{figure}
\includegraphics[width=2.75in,angle=0.]{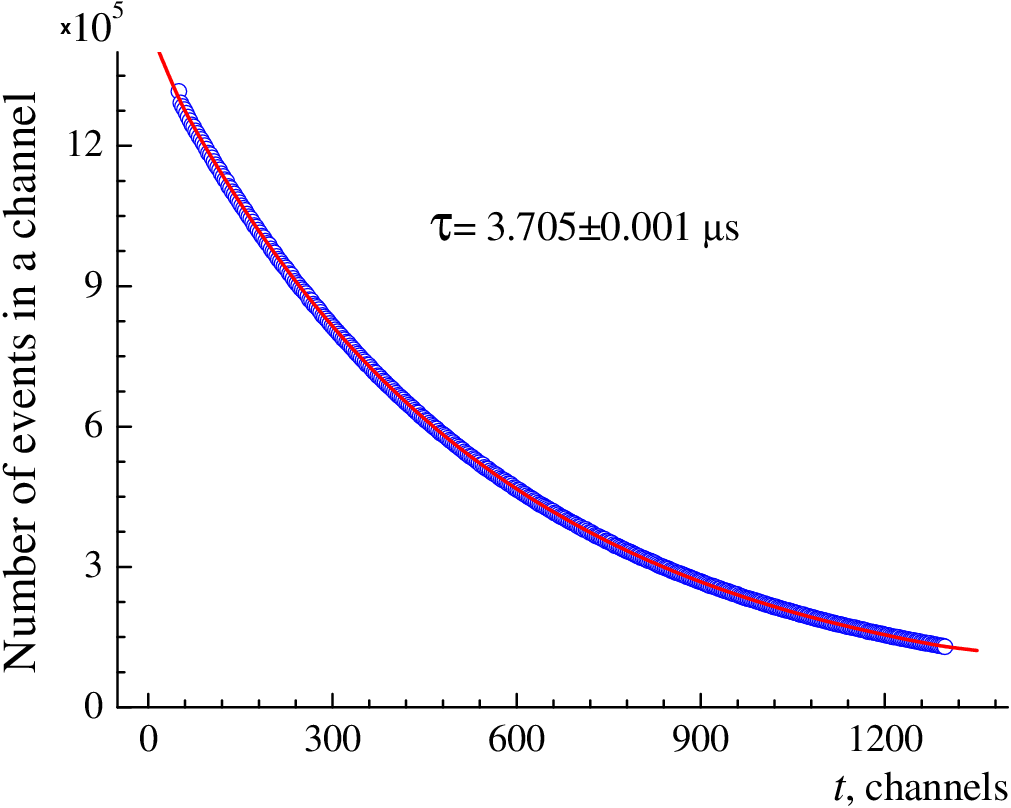}%
\caption{\label{f3} Decay curve for $^{213}$Po plotted by the data from
TAU-3 device obtained over 622 days.}
\end{figure}

Continuous measurements started at TAU-3 on
July 9, 2015. The statistics for 622 days (March 2017) are
processed. In Fig.~\ref{f3}, the decay curve of the $^{213}$Po isotope
is given. The value of $\tau$ was obtained approximating the
decay curve by function $F(t) = A \times exp[-ln(2)t/\tau] + b$
using the minimum $\chi_2$ test in the delay interval of 0.5-13.0 $\mu$s. It was found that $\tau = 3.705 \pm 0.001$ $\mu$s.

The primary data-consistent summation method
was used to find possible periodic dependencies. This
is the method of the interior moving average: to find
harmonics in a data series an interval is chosen about
0.5 of the expected period, and the required parameter
is determined for this interval; then the interval is
shifted by one step and the procedure repeats.

\begin{figure}
\includegraphics[width=2.75in,angle=0.]{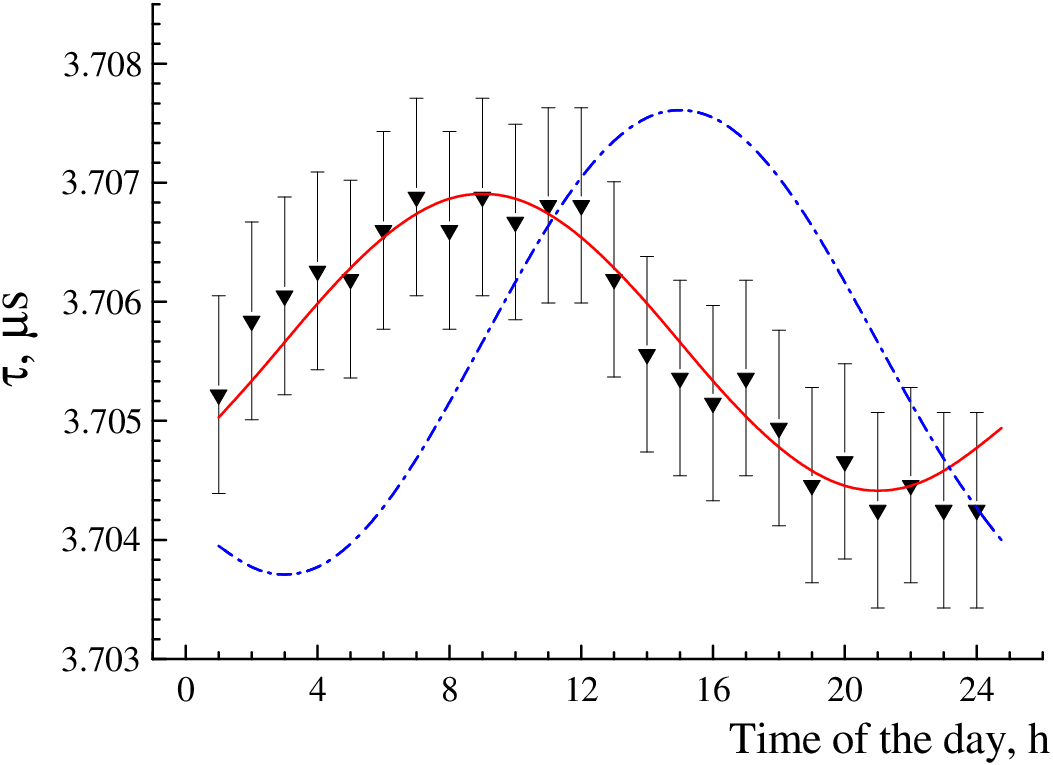}%
\caption{\label{f4} Dependence of $^{213}$Po half-life on the time of solar day obtained by the method of interior moving average
(triangles). Approximation by function $\tau(t) =
\tau_0 [1 + 3.4 \times 10^{-4}sin\{2\pi/24(t-3)\}]$ (red curve).
 Restored dependence $\tau(t) = \tau_0[1 + 5.3 \times 10^{-4}sin\{2\pi/24(t-9)\}]$ (blue dot-dashed curve).}
\end{figure}

In the studies of daily variations of the $^{213}$Po half-life
depending on solar, sidereal, and lunar time, the
length of the respective day was divided into 24 hours.
The length of a sidereal and of a lunar day in the standard
solar time is 23 hours 56 minutes 4.09 s and
24 hours 50 minutes 28.2 s, respectively. A period of
12 hours was chosen as an interval of averaging. The
analysis of events was made as follows. We selected the
events registered in the interval of 0-12 hours for the
entire period study and determined the half-life values.
After that, the interval was shifted by one hour
and the procedure repeated. The results of the search
of the daily variation in solar time are given in Fig.~\ref{f4}.
\begin{figure}
\includegraphics[width=2.8in,angle=0.]{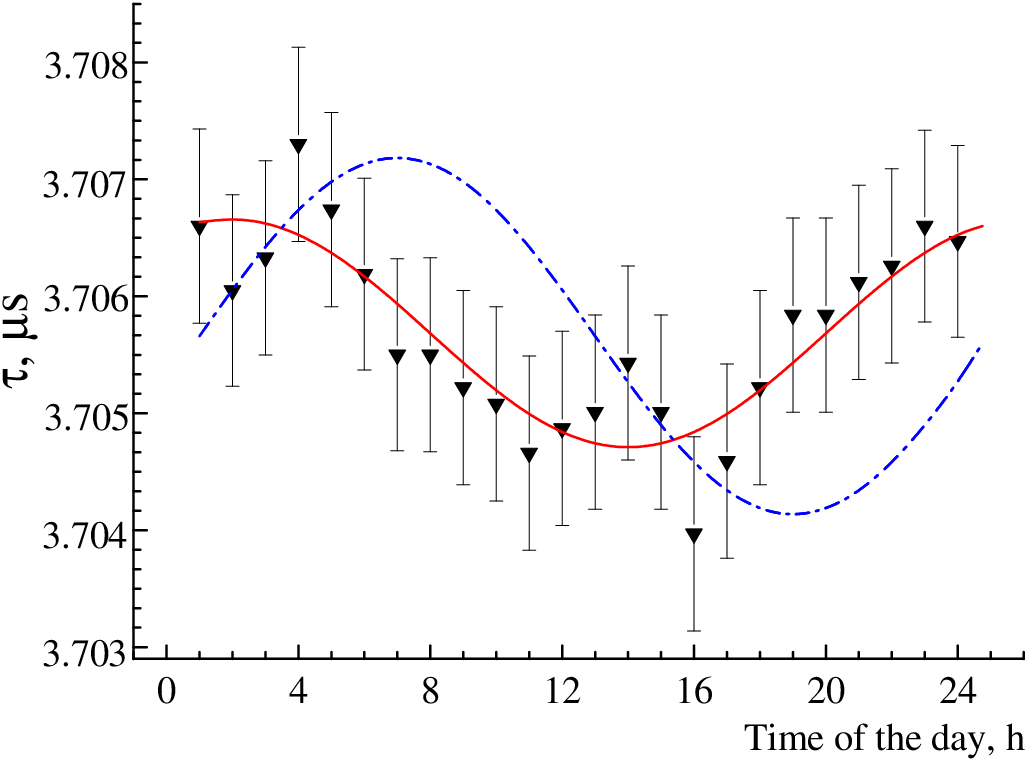}%
\caption{\label{f5} Dependence of $^{213}$Po half-life on the time of sidereal day obtained by the method of interior moving
average (triangles). Approximation by function $\tau(t) =
\tau_0[1 + 2.7 \times 10^{-4}sin\{(2\pi/24)(t - 19)\}]$ (red curve).
Restored dependence $\tau(t) = \tau_0[1 + 4.2 \times 10^{-4}sin\{2\pi/24(t-1)\}$ (blue dot-dashed curve).}
\end{figure}
Here, the result of approximation of the daily half-life
dependence by the function $\tau(t) = \tau _0[1 + Asin\{\omega(t + \phi)\}]$ (red curve) is displayed, where $\tau_0$ is the mean half-life; $\omega = 2\pi/24$ h$^{-1}$; $A = 3.4 \times 10^{-4}$ is the amplitude; $\phi = -3$ h
is a phase shift of the initial point of the curve relative
to 0 hours. The figure shows that the time dependence of the $^{213}$Po half-life is well described by a sinusoidal
function. The period found is 24 hours and the relative
amplitude is 0.00034 half-lives. It is easy to show that
the initial periodic dependence of time data has the
same period (24 h), the amplitude is higher by the factor
of $\pi/2$ and is shifted by 0.5 of the moving interval
($0.25 \times 24 = 6$ h). The amplitude of the initial daily
periodic dependence obtained from these data in solar
time is $A_{So} = (5.3 \pm 1.1) \times 10^{-4}$ (blue dot-dashed curve).

In Fig.~\ref{f5}, the results of the search for a sidereal daily
variation of the $^{213}$Po half-life are displayed. The
experimental data are approximated by curve $\tau(t) =
\tau_o[1 + Asin\{\omega(t + \phi)\}]$ (red curve) with parameters
$A = 2.7 \times 10^{-4}$ is amplitude; $\phi = -19$ h is the phase
shift of the curve initial point relative to 0 hours.

The analysis of the restored initial dependence
similar to the analysis made for the solar-daily wave
shows the presence of a sidereal-daily wave with relative
amplitude $A_S = (4.2 \pm 1.7) \times 10^{-4}$ (blue dot-dashed
curve).

\begin{figure}
\begin{center}
\includegraphics[width=2.8in,angle=0.]{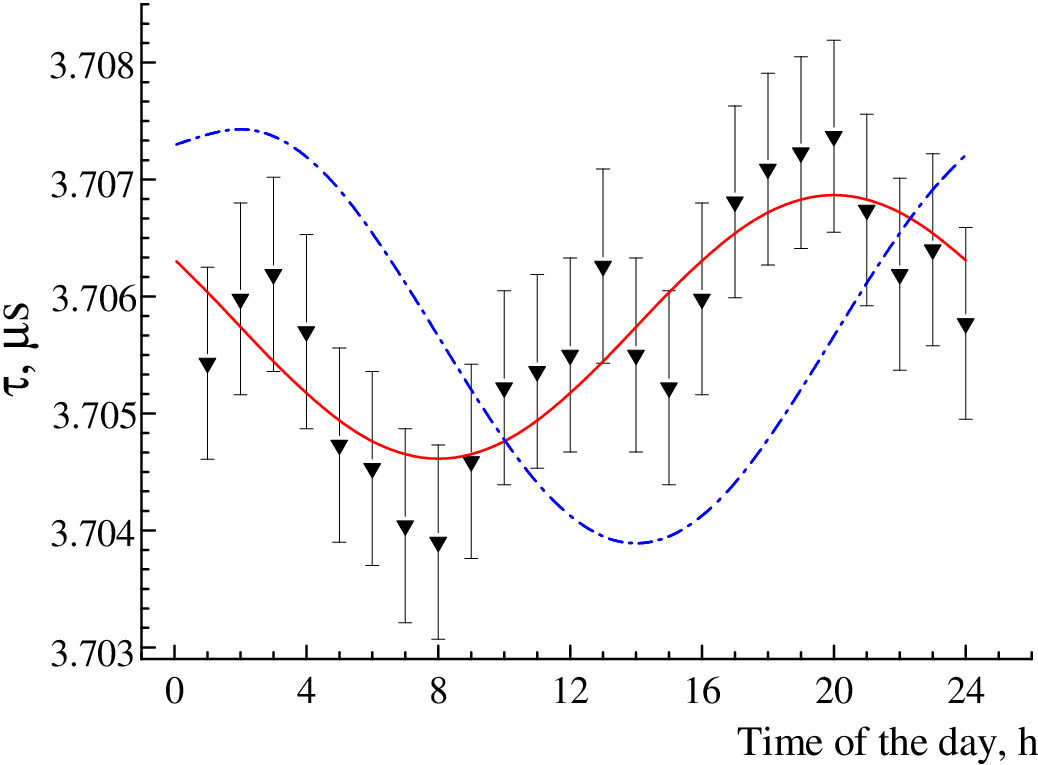}%
\caption{\label{f6} Dependence of $^{213}$Po half-life on the time of lunar day obtained by the method of interior moving average
(triangles). Approximation by function $\tau(t) = \tau_0[1 + 3.1 \times 10^{-4}sin\{2\pi/24(t - 14)\}]$ (red curve).
Restored dependence $\tau(t) = \tau_0[1 + 4.8 \times 10^{-4}sin\{2\pi/24(t -20)\}]$ (blue dot-dashed curve).}
\end{center}
\end{figure}

In Fig.~\ref{f6}, the results of search for a lunar-daily variation
of the $^{213}$Po half-life are given. The analysis of the
restored initial dependence like the analysis made for
the solar-daily wave shows the presence of a lunar-daily
wave with relative amplitude $A_S = (4.8 \pm 2.1) \times 10^{-4}$  (blue dot-dashed curve).

\begin{figure}
\begin{center}
\includegraphics[width=2.8in,angle=0.]{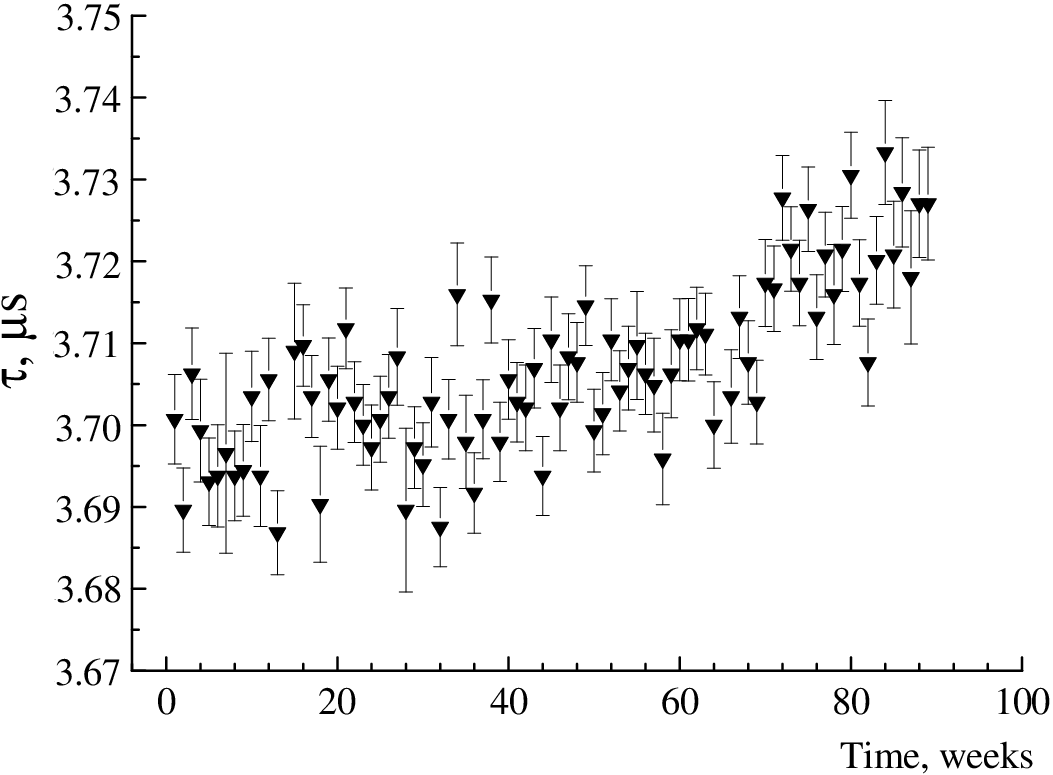}%
\caption{\label{f7} Time dependence of $\tau$ obtained from the decay curve for the weekly data set (start of measurements:
July 9, 2015).}
\end{center}
\end{figure}

\begin{figure}[ht]
\includegraphics[width=2.95in,angle=0.]{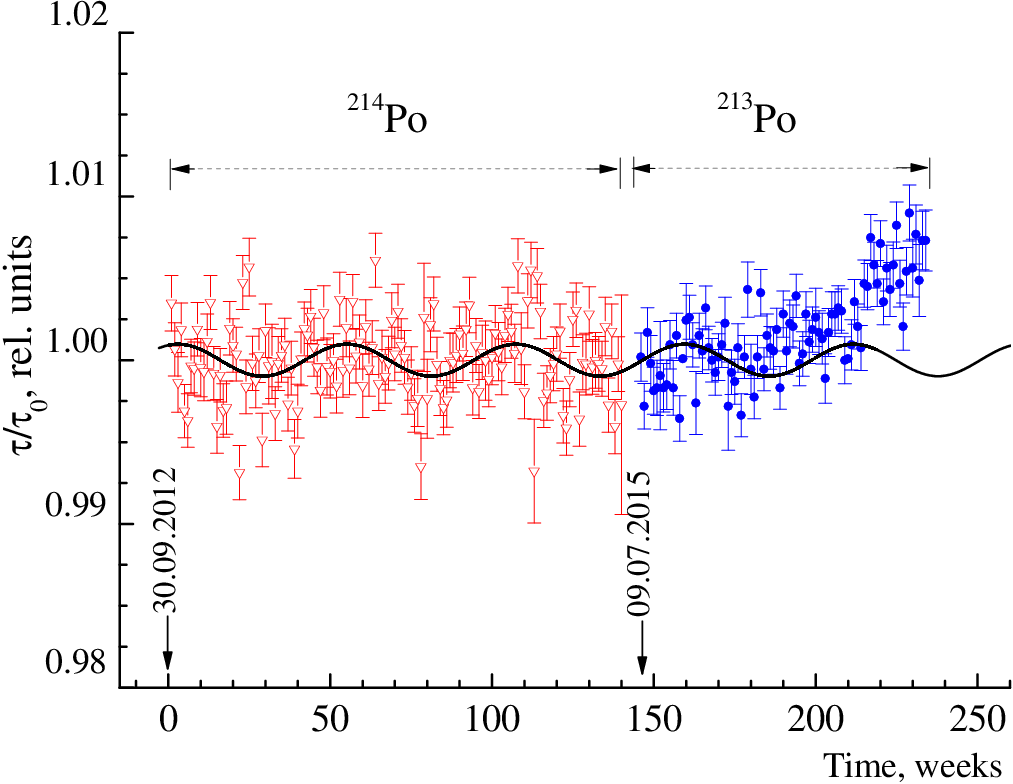}%
\caption{\label{f8}
Superimposed time series of weekly mean normalized
quantities of $^{214}$Po half-life (0-140th week, 973 days)
and $^{213}$Po (146-234$^{\rm th}$ week, 622 days). The sinusoidal function shows an approximation of the annual variation
of $^{214}$Po half-life; it is extrapolated over the entire interval of observations.}

\includegraphics[width=2.75in,angle=0.]{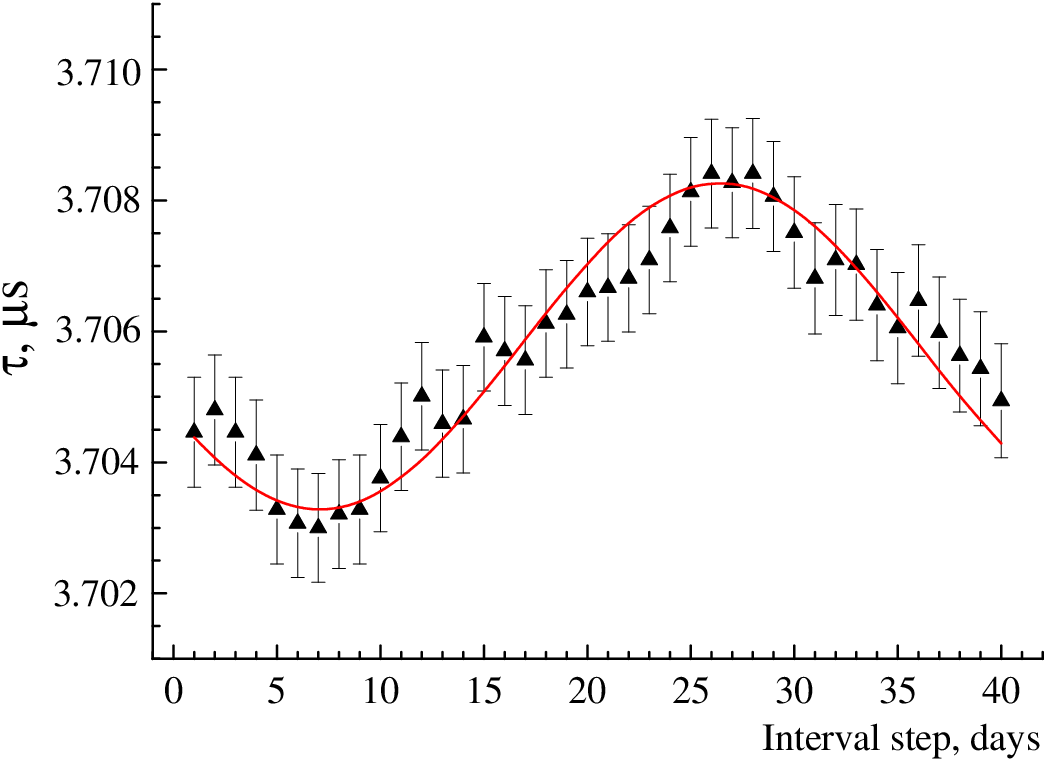}%
\caption{\label{f9}
Time dependence of $^{213}$Po half-life measured
from the beginning of the 38.73-day interval by the method
of interior moving average (triangles) with a one-day step.
Approximation by function $\tau(t) = \tau_0[1 + 6.7 \times
10^{-4}sin\{2\pi/38.73(t - 17.4)\}]$ (red curve).}
\end{figure}

In Fig.~\ref{f7}, the time dependence of $\tau$ obtained from
the decay curve for a weekly data set is presented. It is
shown that $\tau$ increases with time, and that for a data set
collected over 127 days, $\tau = (3.6998 \pm 0.0015)$ $\mu$s; for 320 days, $\tau = (3.6993 \pm 0.0014)$ $\mu$s; for 422 days $\tau=(3.7016 \pm 0.0011)$ $\mu$s, and for 622 days $\tau = (3.7053 \pm 0.0011)$ $\mu$s.
The causes of such behavior of the $\tau$
parameter are not clear yet. It could be both an instrumental
effect, for example, equipment ageing, and an
unknown real physical effect. It seems impossible to
forecast further curve tendency, and only further consistent
measurements can possibly provide the solution
of this problem. The presence of a pulse surge of
data within the time interval comparable to a year in
the series of weekly data hinders using the method of
moving internal average for studies of the half-life
annual variation. Therefore, in order to get a better
understanding of the annual variation, we have
checked a supposition that the half-life annual variation
detected in the series of weekly data on the half-life
of the $^{214}$Po isotope \cite{a3} continues with the same
amplitude and phase in the series of data on $^{213}$Po. The
data normalized to unity for these isotopes in continuous
time are presented in Fig.~\ref{f8}.
The data normalization
for $^{213}$Po was made using the $\tau$ mean value for
320 days. It is shown that the annual variation of the
$^{213}$Po data with the same amplitude and phase as the
data on $^{214}$Po is not excluded. The repeated deviation,
after its shape is specified, can be removed from the
series of data on $^{213}$Po to study the remainder for
annual variations.

\section{DISCUSSION OF RESULTS}

The above-presented results of the $^{213}$Po isotope-decay-constant monitoring for the period of July 2015–
March 2017 show that this parameter was subject to
solar-daily, sidereal-daily, and lunar-daily variations
with amplitudes $A_{So} = (5.3 \pm 1.1) \times 10^{-4}$, $A_S = (4.2 \pm 1.7) \times 10^{-4}$ and $A_L = (4.8 \pm 2.1) \times 10^{-4}$. Within the error, these values coincide with the amplitudes of corresponding variations detected in the series of half-lives of $^{214}$Po. The search for annual variations in $^{213}$Po
data is complicated by the occurrence of aperiodic single-sided deviations of half-lives from the average values
in the time series. The process became visually
noticeable in the section of data recorded over May-June 2016. The effect may be caused by both equipment
ageing and unknown physical factors. To find
the answer we need to continue measurements.

A component with frequency of 9.43 yr$^{-1}$ (period
of 38.73 days) and the maximum power was detected
over the period of June 1996--July 2001 in the analysis
of the power spectrum of frequency components composing
the series of counting rates ($\sim 1/5$ day) at the
Super-Kamiokande facility \cite{a7}. We have searched for
a similar variation in our data series using the method
of the interior moving average. The interval of averaging
was chosen as 19.365 days with a one-day step. The
result of processing is presented in Fig.~\ref{f9} (triangles).
The data were approximated by the function $\tau(t) = \tau_0[1 + Asin\{\omega(t + \phi)\}]$, where $\tau_0$ is the mean half-life;
$t$ is time [days]; $\omega = 2\pi/38.73$ day$^{-1}$; $A = (6.7 \pm 1.1) \times 10^{-4}$ is amplitude; $\phi = -17.4$ day is the phase shift of
the curve initial point relative to zero. The restored
wave amplitude $A = (10.6 \pm 1.7) \times 10^{-4}$ was obtained
from the approximation by multiplying by $\pi/2$. In
order to verify the result stability, a similar procedure
was performed with the data of the TAU-2 facility
accumulated for 590-day measurements with the $^{214}$Po isotope. The restored wave amplitude was $A = (10.6 \pm 1.9) \times 10^{-4}$. The search for a wave with the frequency
of 10 yr$^{-1}$ was carried out for verification. Within the
statistical error of $\pm1.9 \times 10^{-4}$, no variation with this
frequency was detected. Thus, we can conclude that
the variation with the frequency of 9.43 yr$^{-1}$ is of global
character (at least, for the underground devices),
though does not coincide with known natural
rhythms. In \cite{a7}, the authors consider the possibility
that the Sun’s core can feature a similar rhythm, and
study possible mechanisms of these variations.

\section{CONCLUSIONS}

In this study, to search for variations of $^{213}$Po half-life
($T_{1/2} = 3.7$ $\mu$s), a device with $^{229}$Th isotope as a parent
source was constructed. A solar-daily variation
with amplitude $A_{So} = (5.3 \pm 1.1) \times 10^{-4}$, a lunar-daily
variation with amplitude $A_L = (4.8 \pm 2.1) \times 10^{-4}$, and
a sidereal-daily variation with amplitude $A_S = (4.2 \pm 1.7) \times 10^{-4}$ were discovered upon processing the data
series for the period from July 2015 to March 2017
(622 days). The $^{213}$Po half-life value averaged over 662 days is found to be $T_{1/2} = 3.705 \pm 0.001$ $\mu$s.
This is consistent with the result ($T_{1/2} = 3.708 \pm 0.008$ $\mu$s) obtained by means of an ion-implanted planar Si detector for alpha and beta particles emitted from weak $^{225}$Ac sources in work \cite{a4b}.

\begin{center}
\textbf{ACKNOWLEDGMENTS}
\end{center}

The study was supported by the High Energy Physics
and Neutrino Astrophysics Program of the Presidium
of the Russian Academy of Sciences.

\end{document}